\documentclass[preprint,aps,preprintnumbers,nofootinbib,showpacs]{revtex4-1}
\pdfoutput=1
\usepackage{graphicx,color,amsmath,amsfonts,multirow}
\usepackage{epsfig}
\usepackage{rotating}
\usepackage{array}
\usepackage{graphicx}
\usepackage{multirow}
\usepackage{epstopdf}
\usepackage{amssymb}
\usepackage{latexsym}
\newcommand{\fig}{\begin{figure}}
\newcommand{\ef}{\end{figure}}

\begin{document}
\draft

\topmargin=-1.5cm

\title{\mbox{}\\[10pt]
 Prediction of  leptonic CP phase from\\
 perturbatively modified tri-bimaximal (or bimaximal) mixing }

\author{Sin Kyu Kang$^{a,b}$\footnote{E-mail:
        skkang@seoultech.ac.kr},
C. S. Kim$^c$\footnote{E-mail: cskim@yonsei.ac.kr}}
\affiliation{$^a$~School of Liberal Arts, Seoul-Tech, Seoul 139-743, Korea}
\affiliation{$^b$~PITT PACC, Dept. of Physics and Astronomy, Univ. of Pittsburgh, Pittsburgh, PA 15260, USA}
\affiliation{$^c$Dept. of Physics and IPAP, Yonsei Univ., Seoul 120-749, Korea}

%\thanks{}


\date{\today}

%%%%%%%%%%%%%%%%%%%%%%%%%%%%%%%%%%%%%%%%%%%%%%%%%%%%%%%%%%%%%%%%%%%%%%%%%%%%%%%%%%%%%%%%%%
\begin{abstract}
\noindent We consider the perturbatively modified tri-bimaximal (or bimaximal) mixing
to estimate the (Dirac-type) CP phase in the neutrino mixing matrix.
The expressions for the CP phase are derived from the equivalence  between the standard parametrization of the neutrino mixing matrix for the Majorana neutrino
and modified tri-bimaximal or bimaximal mixing matrices with appropriate CP phases.
Carrying out numerical analysis based on the current experimental results for neutrino mixing angles, we can predict the values of the CP phase for several possible cases.

\noindent

\end{abstract}
%\pacs{14.60.Pq, 14.60.Lm,11.30.Er}

\maketitle

%%%%%%%%%%%%%%%%%%%%%%%%%%%%%%%%%%%%%%%%%%%%%%%%%%%%%%%%%%%%%%%%%%%%%%%%%%%%%%%%%%%%%%%%%%\section

%\newpage
%%%%%%%%%%%%%%%%%%%%%%%%%%%%%%%%%%%%%%%%%%%%%%%%%%%%%%%%%%%%%%%%%%%%%%%%%%%%%%%%%%%%%%%%%%
\section{ Introduction}

Recent measurements of not-so-small value of the reactor neutrino mixing angle
have opened up new windows to probe leptonic CP violation (LCPV) \cite{theta13}.
Establishing LCPV is one of the most challenging tasks in future neutrino experiments \cite{cpv-lepton}.
The  PMNS neutrino mixing matrix \cite{pmns}  is presented by  $3\times 3$ unitary matrix which contains, in addition to the three angles, a Dirac type CP violating phase in general as it exists in the quark sector, and two extra phases if neutrinos are Majorana particles.
Although we do not yet have compelling evidence for LCPV, the current fit to neutrino data indicates nontrivial values of the Dirac-type CP phase  \cite{cp-fit1,cp-fit2}.
Several experiments have been proposed or being scheduled  to establish CP violation in neutrino oscillations
\cite{cpv-exp}.
In this situation, it must deserve to investigate possible size of LCPV detectable through neutrino oscillations.
From the point of view of {\it calculability}, it is conceivable that a Dirac type LCPV phase may be calculable with regards to some observables  \cite{cp-angle}. In this letter, we propose possible forms of neutrino mixing matrix that lead us to estimate the possible size of LCPV phase, particularly, in terms of
 two neutrino mixing angles {\it only}, in the PDG-type standard parametrization \cite{pdg}.
The estimation of LCPV phase is carried out  by the following procedures:
\begin{itemize}
\item{ Constructing the neutrino mixing matrix with appropriate CP phases so as to accommodate  the current neutrino oscillation data in such a way to perturb  conventional (tri-)bimaximal matrix.}
\item{ Deriving  the master formulae
linking the Dirac-type CP phase with neutrino mixing angles from the equivalence principle that any forms of neutrino mixing matrix should be equivalent to the standard parametrization of the PMNS mixing matrix.}
\end{itemize}
As will be shown later, the neutrino mixing matrices we adopt at the first step contain a maximal mixing angle which  plays a crucial role in deriving the relations among  neutrino mixing angles and Dirac-type CP phase
in the standard parametrization.
Substituting  values of  neutrino mixing angles into those equations obtained at the second step, we perform numerical analysis on observables for the LCPV and present the results.

%%%%%%%%%%%%%%%%%%%%%%%%%%%%%%%%%%%%%%%%%%%%%%%%%%%%%%%%%%%%
\section{ Neutrino Mixing Matrices}

 In the leading order approximation, the conventional neutrino mixing matrices in the flavor basis can be given by
\begin{eqnarray}
 U_0^{\rm PMNS} =
 \left(\begin{array}{ccc}
 1 & 0 & 0 \\
 0 &  \frac{1}{\sqrt{2}}&  -\frac{1}{\sqrt{2}} \\
 0 &  \frac{1}{\sqrt{2}} &  \frac{1}{\sqrt{2}}
 \end{array}\right)
 \left(\begin{array}{ccc}
  \cos\theta & -\sin\theta & 0\\
 \sin\theta &  \cos\theta &  0 \\
 0 &  0 &  1
 \end{array}\right). \label{leading-mixing}
 \end{eqnarray}
Taking $\sin\theta$ to be either $1/\sqrt{3}$ or $1/\sqrt{2}$ leads to the so-called tri-bimaximal mixing $U_{0}^{\rm TBM}$  or bimaximal mixing $U_{0}^{\rm BM}$,  respectively \cite{tribi,bimax}
Although the tri-bimaximal and/or bimaximal ones are theoretically well-motivated patterns of the neutrino mixing matrix,
they are challenged by the current  experimental results for three neutrino mixing angles.
While the bimaximal mixing has already been ruled out by the non-maximal mixing for the solar angle, the current measurements of non-zero  $\theta_{13}$ definitely disfavor the exact tri-bimaximal mixing.
Since the measured values of $\theta_{13}$ have been turned out to be of order of the required deviation of $\theta_{12}$ from maximal,  the tri-bimaximal mixing can be treated on the same footing with the bimaximal mixing as leading order approximation of the neutrino mixing matrix.

The simplest possible forms of the neutrino mixing matrix without CP phases deviated from the  (tri-)bimaximal  mixing patterns are given by \cite{one-unitary,cases}
\begin{eqnarray}
\left\{ \begin{array}{l}
   U_{0}^{\rm (T)BM} \cdot U_{ij}(\theta) ,\\
  U^{\dagger}_{ij}(\theta) \cdot U_{0}^{\rm (T)BM},\\
     \end{array} \right.
    \label{phno1}
    \end{eqnarray}
where  $U_{ij}(\theta)$ represents the unitary matrix corresponding to the rotation with
the angle $\theta$ in $(i,j)$ plane.
Please note that $U_0^{\rm (T)BM}$ can be achieved  in a neutrino model
with a flavor symmetry by breaking it down to two different residual symmetries preserved in the neutrino and the charged lepton sector respectively \cite{flavor-sym}. In such a model, $U_{ij}(\theta)$ in the upper (lower) form of Eq.(\ref{phno1}) is arisen from appropriate breaking of the residual symmetry of the mass matrix in the neutrino (charged lepton) sector by adding a bearking term in (i,j) and (j,i) entries of the mass matrix.
Once the mixing angle $\theta$ can perturbatively be  treated \cite{one-unitary}, then Eq. (2) possibly  gives rise to non-zero value of the reactor angle as well as deviation from the maximal for the solar angle.
As will be shown later, eight forms among twelve possible ones in Eq. (\ref{phno1}) are in consistent with present neutrino data within $3\sigma$ C.L.
In this respect, we  call those eight forms of the neutrino mixing matrix {\it  minimally-modified (tri-)bimaximal} (M(T)BM) parameterizations.
It is worthwhile to notice that  those forms of the neutrino mixing matrix keep a column or a row  in  (tri-)bimaximal  mixing matrix  unchanged, which may be regarded as a remnant of a possible horizontal symmetry leading to (tri-)bimaximal mixing.
The column vectors orthogonal to the $i$-th and $j$-th ones in
$U_{0}^{\rm (T)BM}$ are unchanged
for $U_{0}^{\rm (T)BM}U_{ij}(\theta)$, whereas the row vectors orthogonal to the $i$-th and $j$-th ones are unchanged for   $U^{\dagger}_{ij}(\theta)U_{0}^{\rm (T)BM}$.
The multiplication of  $U_{ij}(\theta)$ represents unitary transformation of the symmetry operator which corresponds to the rotation of  two column vectors in the mixing matrix.
Thus, a symmetry argument\footnote{C. S. Lam has shown \cite{cslam} that the column vectors of the lepton mixing matrix can be eigenvectors of certain horizontal flavor symmetry.
Keeping the symmetry point of view,  we can construct the M(T)BM parameterizations of the neutrino mixing matrix  by appropriately multiplying
either tri-bimaximal or bimaximal mixing matrix by a unitary matrix while keeping a column or a row vector characterizing the horizontal symmetry unchanged.
}
can still be applied to the origin of  the neutrino mixing matrices in the M(T)BM parameterizations.

Since the Dirac-type CP phase $\delta_D$ is accompanied by $\theta_{13}$ in the standard parametrization, it is natural to involve CP phases when construct  neutrino mixing matrix so as to generate non-zero $\theta_{13}$.
Interesting points in this work based on the simplest forms of neutrino mixing matrix aforementioned are that  $\theta_{13}$ is related with either $\theta_{12}$ or $\theta_{23}$, and $\delta_D$ 
can be related in the standard parametrization with two neutrino mixing angles as long as we identify
the M(T)BM parameterizations with the standard one.
%In this work,
Therefore, It is highly desirable to predict the Dirac-type CP phase with complex perturbations $U_{ij}(\theta,\xi)$ containing a phase $\xi$.
Among the above twelve forms of the mixing matrix given in Eq. (\ref{phno1}), the forms $ U_{0}^{\rm (T)BM}U_{12}(\theta, \xi)$ and $U^{\dagger}_{23}(\theta, \xi) U_{0}^{\rm (T)BM}$ still lead to vanishing reactor mixing angle, and thus predict no CP violation. We do not consider these cases any longer.
Therefore, all the possible forms of the MT(B)M mixing matrix eligible for our aim are presented as follows;
%%%%%%%%%%%%%%%%%%%%%%%%%%%%%%%%%%%%%%%%%%%%%%%%%%%%%%
 \begin{eqnarray}
 V=\left\{ \begin{array}{l}
     U_{0}^{\rm TBM} U_{23}(\theta, \xi) ~~~\mbox{(Case--A)},\\
     U_{0}^{\rm TBM} U_{13}(\theta, \xi) ~~~\mbox{(Case--B)},\\
     U^{\dagger}_{12}(\theta, \xi) U_{0}^{\rm TBM}  ~~~\mbox{(Case--C)},\\
     U^{\dagger}_{13}(\theta, \xi) U_{0}^{\rm TBM}  ~~~\mbox{(Case--D)}, \\
    U^{\dagger}_{12}(\theta, \xi) U_{0}^{\rm BM}  ~~~~\mbox{(Case--E)},\\
     U^{\dagger}_{13}(\theta, \xi) U_{0}^{\rm BM}  ~~~~\mbox{(Case--F)},\\
    U_{0}^{\rm BM} U_{23}(\theta, \xi) ~~~~\mbox{(Case--G)},\\
     U_{0}^{\rm BM} U_{13}(\theta, \xi) ~~~~\mbox{(Case--H)}.
     \end{array}\right.
    \label{phno3}
    \end{eqnarray}
While the cases (A) , (C) and (E) have been studied in \cite{cases}, the other cases have not been considered yet.
For the completeness of possibility,
 we here propose a general way to extract Dirac-type CP phase  from all possible forms given in Eq.(\ref{phno3}), and show that  not only the cases (A), (C), (E)
 but also the other cases (B), (D), (F) are still viable from the recent fit of neutrino mixing angles up to $3\sigma$ C.L. \cite{cp-fit2}.
%We also investigate how different the values of CP phase are predicted in each case.

%
%
\section{ Calculation of Leptonic CP violation }

Now we demonstrate how to derive $\delta_D$  in terms of neutrino mixing angles
in the standard parametrization. This can be done from the equivalence between  one of the parameterizations  in Eq. (\ref{phno3})
and the standard parametrization, shown in Eq. (\ref{standard}).

Assuming that neutrinos are Majorana particles, we begin by explicitly presenting the  PMNS neutrino mixing matrix in the PDG-type standard parametrization as follows \cite{pdg},
\begin{eqnarray}
&& U^{\rm ST} = U^{\rm PMNS}\cdot P_{\phi}\nonumber \\
 &&=
 \left(\begin{array}{ccc}
 c_{12}c_{13} & s_{12}c_{13} & s_{13}e^{-i\delta_D}\\
 -s_{12}c_{23}-c_{12}s_{23}s_{13}e^{i\delta_D} &  c_{12}c_{23}-s_{12}s_{23}s_{13}e^{i\delta_D}&  s_{23}c_{13} \\
 s_{12}s_{23}-c_{12}c_{23}s_{13}e^{i\delta_D} &  -c_{12}s_{23}-s_{12}c_{23}s_{13}e^{i\delta_D}&  c_{23}c_{13} \\
 \end{array}\right)P_{\phi} ,
 \label{standard}
 \end{eqnarray}
 where $P_{\phi} \equiv {\rm Diag.}(e^{i\phi_1}, e^{i\phi_2},e^{i\phi_3})$ is a $3\times 3$ phase matrix.
Note  that  one of three phases in $P_{\phi}$ is redundant.
 Incorporating phase matrices $P$ defined above,  the neutrino mixing matrices in Eq. (\ref{phno3}) are given by
 $$U^{\rm ST} = P_{\alpha} \cdot V \cdot P_{\beta}.$$
Without those two phase matrices $P_{\alpha}$ and $P_{\beta}$, in general, we cannot equate the M(T)BM parameterizations with the standard parametrization  given in Eq. (\ref{standard}).
 Please note that such bi-unitary transformation is regarded as a general basis change of leptonic fields \cite{Antusch}.
The equivalence between both parameterizations dictates the following relation,
\begin{eqnarray}
V_{ij}e^{i(\alpha_i + \beta_j)} = U^{\rm ST}_{ij}
= U^{\rm PMNS}_{ij} e^{i\phi_j} ~. \label{rel1}
\end{eqnarray}
Applying $|V_{13}|=|U_{13}^{\rm ST}|$ and $ |V_{11}/V_{ 12}|=|U_{11}^{\rm ST}/U_{12}^{\rm ST}|$ to Cases A, B, G and H, we obtain the relations between
the solar and  reactor mixing angles,
\begin{eqnarray}
 s^2_{12}=\left\{ \begin{array}{r}
   1-\frac{2}{3(1-s^2_{13})}~~~\mbox{(Case--A)},\\
     \frac{1}{3(1-s^2_{13})} ~~~\mbox{(Case--B)}, \\
     1-\frac{1}{2(1-s^2_{13})}~~~\mbox{(Case--G)},\\
     \frac{1}{2(1-s^2_{13})} ~~~\mbox{(Case--H)}. \\
     \end{array}\right.
    \label{angle1}
\end{eqnarray}
Those relations indicate that non-zero values of $s^2_{13}$ lead to  $s^2_{12}<1/3$ for Case--A and $s^2_{12}>1/3$ for Case--B.
While the results for Case--A  are well consistent with the current experimental values of $s_{12}^2$  at $1\sigma$ C.L., those for Case--B are so at $2\sigma$ C.L.
It turns out that the above relations for Cases G and H are not  consistent with experimental results up to $3 \sigma$ C.L., and thus ruled out.

Similarly,  we get the relations between
the atmospheric and reactor mixing angles from  $|V_{13}|=|U_{13}^{\rm ST}|$ and $ |V_{23}/V_{ 33}|=|U_{23}^{\rm ST}/U_{33}^{\rm ST}|$,
\begin{eqnarray}
 s^2_{23}=\left\{ \begin{array}{r}
    1-\frac{1}{2(1-s^2_{13})}~~~\mbox{(Cases--C and --E)},\\
    \frac{1}{2(1-s^2_{13})} ~~~\mbox{(Cases--D and --F)}.
     \end{array}\right.
    \label{angle2}
\end{eqnarray}
We see that non-zero values of $s_{13}^2$ lead to  the values of  $s_{23}^2<1/2$  for Cases--C and --E  and $s_{23}^2>1/2$ for Cases--D and-- F.
%For the experimental values of $s_{13}^2$ at $1\sigma$ C.L.,  the values of $s_{23}^2$ are estimated to be around 0.48 (0.511) for Case--C (D). which
They turned out  to be consistent with experimental values of $s_{23}^2$ at 2$\sigma$ C.L.

%%%%%%%%%%%%%%%%%%%%%%%%%%%%%%%%%%%%%%%%%
\begin{table}[b]
\caption{Formulae for $\cos\delta_{D}$ and ${\rm J_{CP}^2}$ for Cases  B -- F. The second column corresponds to the relation (\ref{relA-3}) for Case A. $\eta_{ij}=\frac{1}{2\tan 2 \theta_{ij}}$,
$\kappa_{ij}=\cos^2 2\theta_{ij}\cdot c_{13}^4$, $\xi= \sin 2 \theta_{12}$ and
$\omega=(s^2_{13} (9s^2_{12}-4)-3 s^2_{12} +1 )^2$}
\label{tab:table1}
\begin{center}
\begin{tabular}{|c|c|c|c|}
               \hline \hline
          Cases &   & $\cos\delta_D$ & $ {\rm J^2_{CP}}$ \\
\hline\hline
            B & $ \frac{V_{21}+V_{31}}{V_{23}+V_{33}}=\frac{V_{11}}{V_{13}}$ &
                  $\frac{2-4 s^2_{13}}{s_{13} \sqrt{2-3 s^2_{13}}} \eta_{23}$ &
                           $\frac{1}{6^2}[s^2_{13}(2-3s^2_{13})-\kappa_{23}]$ \\
\hline
            C & $ \frac{V_{11}+\sqrt{2} V_{12}}{V_{21}+\sqrt{2} V_{22}}=\frac{V_{13}}{V_{23}}$ &$\frac{s^2_{13}-(1-3 s^2_{12})(1-3 s^2_{13})}{3s_{13} \sqrt{1-2 s^2_{13}}\xi} $&
                           $\frac{1}{12^2}[9c^2_{12} s^2_{12}s^2_{13}(1-2s^2_{13})-\omega]$ \\
\hline
            D & $ \frac{V_{11}+\sqrt{2} V_{12}}{V_{31}+\sqrt{2} V_{32}}=\frac{V_{13}}{V_{23}}$ &$\frac{(1-3s^2_{13})(1-3s^2_{12})-s^2_{13}}{3 s_{13} \sqrt{1-2 s^2_{13}}\xi}  $&
                           $\frac{1}{12^2}[9c^2_{12} s^2_{12}s^2_{13}(1-2s^2_{13})-\omega]$ \\
\hline
              E & $ \frac{V_{12}+V_{11}}{V_{21}+V_{22}}=\frac{V_{13}}{V_{23}}$ &$-\frac{1-3 s^2_{13}}{s_{13} \sqrt{1-2 s^2_{13}}} \eta_{12} $&
                           $\frac{1}{8^2}[4s^2_{13}(1-2s^2_{13})-\kappa_{12}]$ \\
\hline
            F & $ \frac{V_{11}+V_{12}}{V_{32}+V_{31}}=\frac{V_{13}}{V_{33}}$ &$\frac{1-3 s^2_{13}}{s_{13} \sqrt{1-2 s^2_{13}}} \eta_{12} $&
                           $\frac{1}{8^2}[4s^2_{13}(1-2s^2_{13})-\kappa_{12}]$ \\
               \hline \hline
\end{tabular}
\end{center}
\end{table}

Now, let us derive the relations among $\delta_D$ and neutrino mixing angles in the standard parametrization.
Since the same method can be applied to all the cases, we only present how to derive the relation only for Case--A.
From the components of the neutrino mixing matrix  for Case--A, we see that
\begin{eqnarray}
  \frac{V_{23}+V_{33}}{V_{22}+V_{32}}=\frac{V_{13}}{V_{12}}. \label{relA-3}
  \end{eqnarray}
Applying the relation (\ref{rel1}) and  $V_{21}=V_{31}$ to Eq. (\ref{relA-3}), we can get the relation
  \begin{eqnarray}
  \frac{U^{\rm ST}_{13}}{U^{\rm ST}_{12}}=\frac{U^{\rm ST}_{23}U^{\rm ST}_{31}+U^{\rm ST}_{33}U^{\rm ST}_{21}}{U^{\rm ST}_{22}U^{\rm ST}_{31}+U^{\rm ST}_{32}U^{\rm ST}_{21}}. \label{A-final}
  \end{eqnarray}
 Presenting $U^{\rm ST}_{ij}$ in terms of  the neutrino mixing angles as well as  $\delta_D$,
and taking the real part in Eq. (\ref{A-final}), we get the equation for $\delta_D$ as
 \begin{eqnarray}
 \cos\delta_D=\frac{-1}{2\tan 2 \theta_{23}}\cdot \frac{1-5 s^2_{13}}{s_{13} \sqrt{2-6 s^2_{13}}}. \label{phaseA}
 \end{eqnarray}
 Notice that the imaginary part in Eq. (\ref{A-final}) is automatically cancelled.
Using the above formulae, we can easily derive the leptonic Jarlskog invariant as follows;
\begin{eqnarray}
{\rm J^2_{CP}}&=& ({\rm Im}[U^{\rm ST}_{\mu 2}U^{\rm ST}_{e 3}U^{\rm ST\ast}_{e2} U^{\rm ST \ast}_{\mu 3}])^2
                  \nonumber \\
&=& \frac{1}{8} \sin (2\theta_{12})\sin(2\theta_{13})\sin(2\theta_{23})\sin\delta_D \label{jak} \\
          &=& \frac{1}{12^2}[8s^2_{13}(1-3s^2_{13})-\cos^2 2\theta_{23} c_{13}^4], \label{jcp1}
\end{eqnarray}
where Eq.(\ref{jak}) is obtained \cite{cpv-lepton} by just inserting  the entries of $U^{\rm ST}_{\alpha i}$ given in
Eq.(\ref{standard}).

By taking the same procedure described above, we can obtain the formulae for $\delta_D$ and ${\rm J}^2_{CP}$ for Cases B -- F as presented in
 in Table I. Note that the Cases G and H are experimentally ruled out as previously mentioned.
\\

%
%We see that the Jarlskog invariants depend on only $\theta_{13}$ and $\theta_{23}$ for Cases A and B,
% and $\theta_{13}$ and $\theta_{12}$ for Cases C and D.
%Note that  ${\rm J^2_{CP}}$ for Cases C and D are the same.
%%%%%%%%%%%%%%%%%%%%%%%%%%%%%%%%%%%%%%%%%%%%%%%%%%%%%%%%%%%%%%%%%%%%%%%

\begin{figure}[h]
\begin{center}
\includegraphics[width=0.49\linewidth]{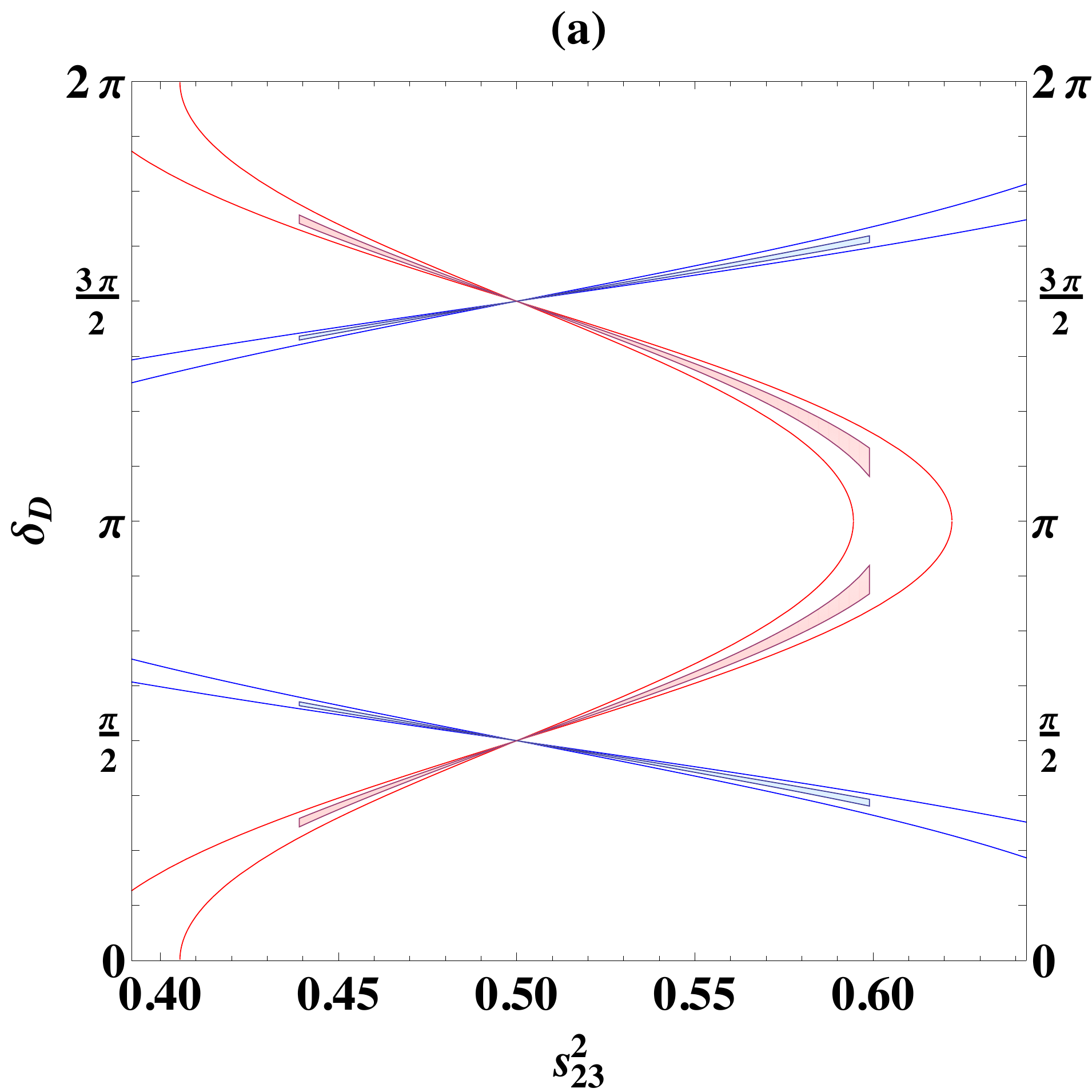}
\includegraphics[width=0.49\linewidth]{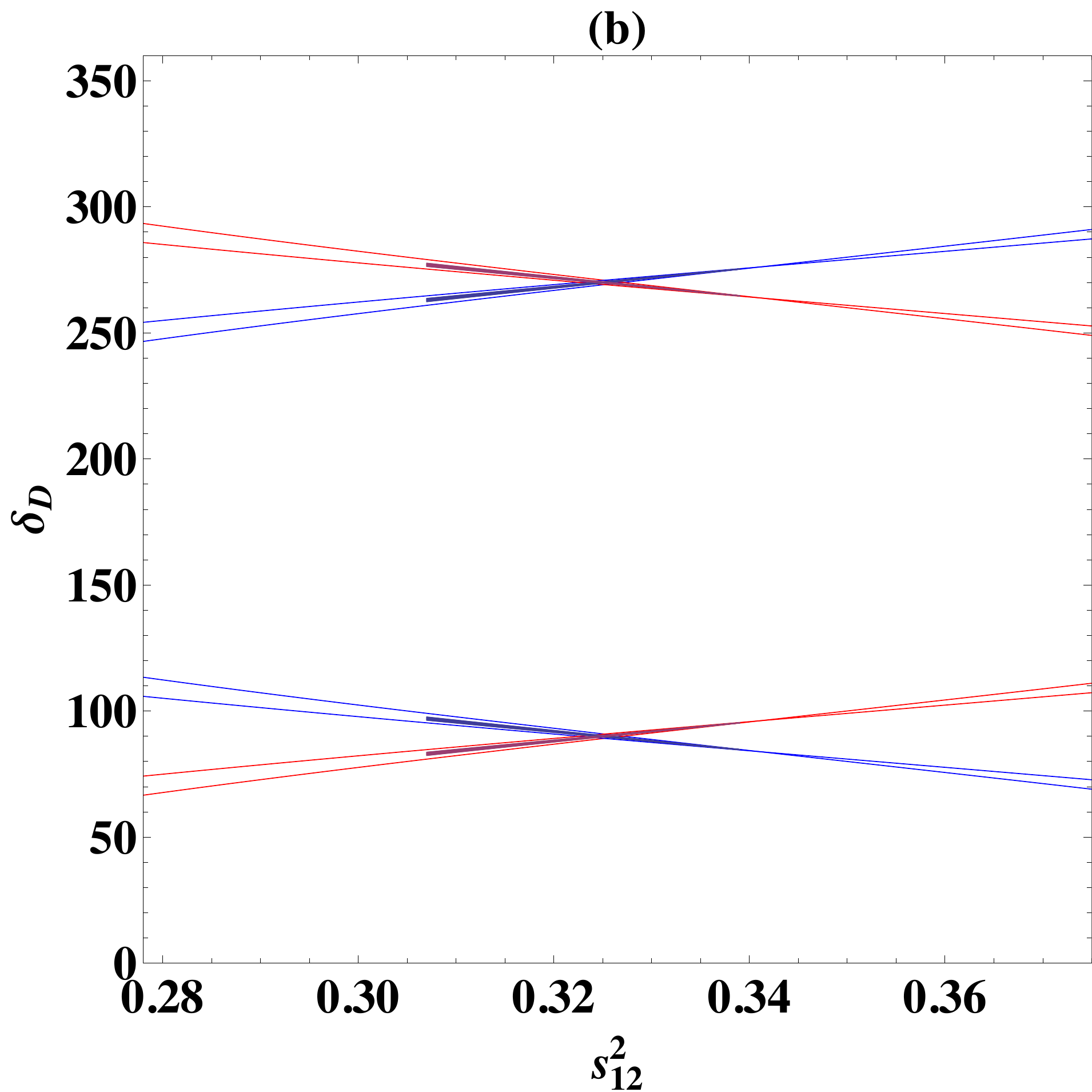}
\includegraphics[width=0.49\linewidth]{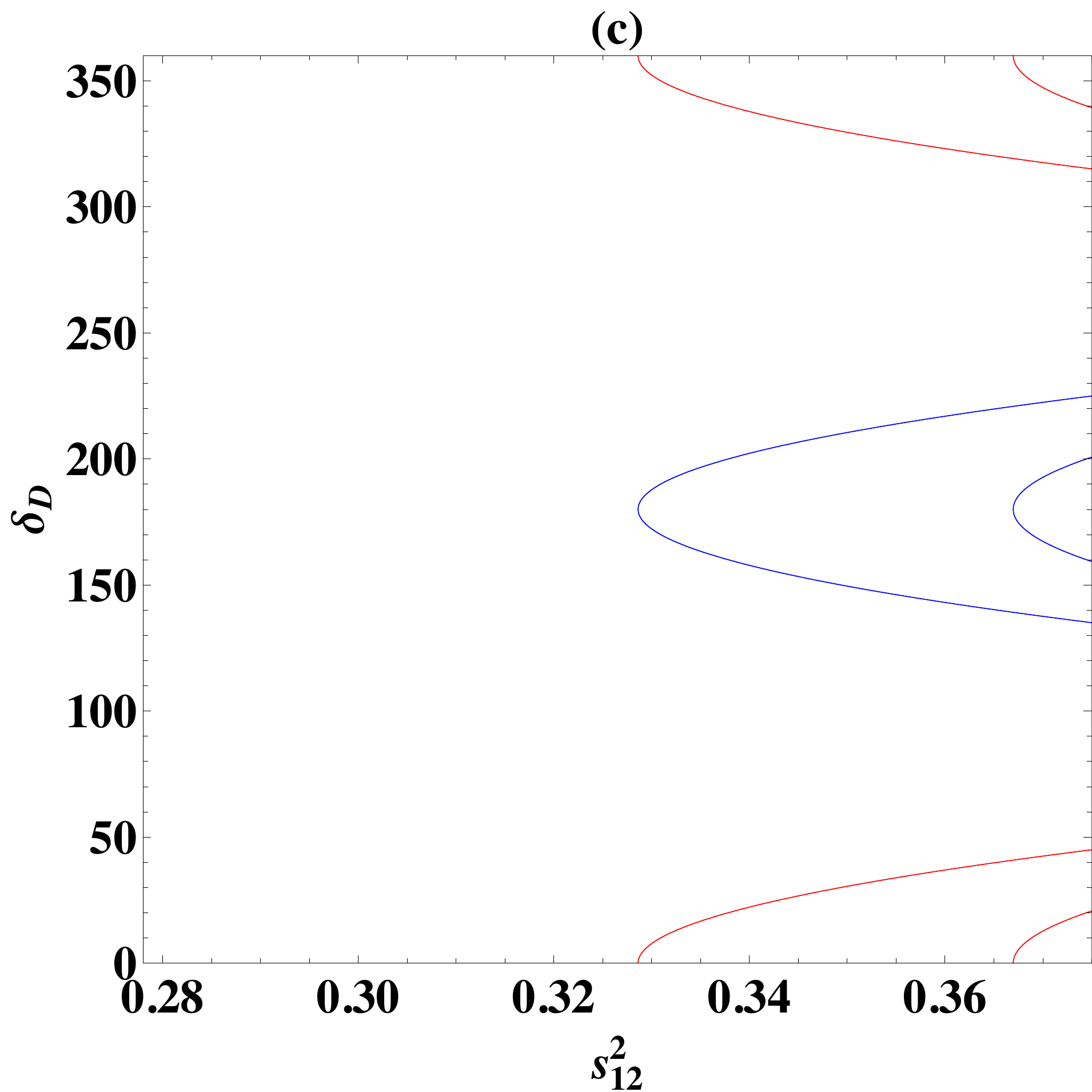}
\end{center}
\caption{Predictions of $\delta_D$ in terms of $s_{23}^2$ [(a): Cases A and B] and $s_{12}^2$ [(b): Cases C  and D] [(c):Cases E and F]  based on the experimental data  at $3\sigma$ and $1\sigma$  ( for Cases A-D)
C.L.
Regions surrounded by  blue (red) lines correspond to Cases A, C and E (B, D and F).
}
\label{fig1}
\end{figure}

\begin{figure}[h]
\begin{center}
\includegraphics[width=0.6\linewidth]{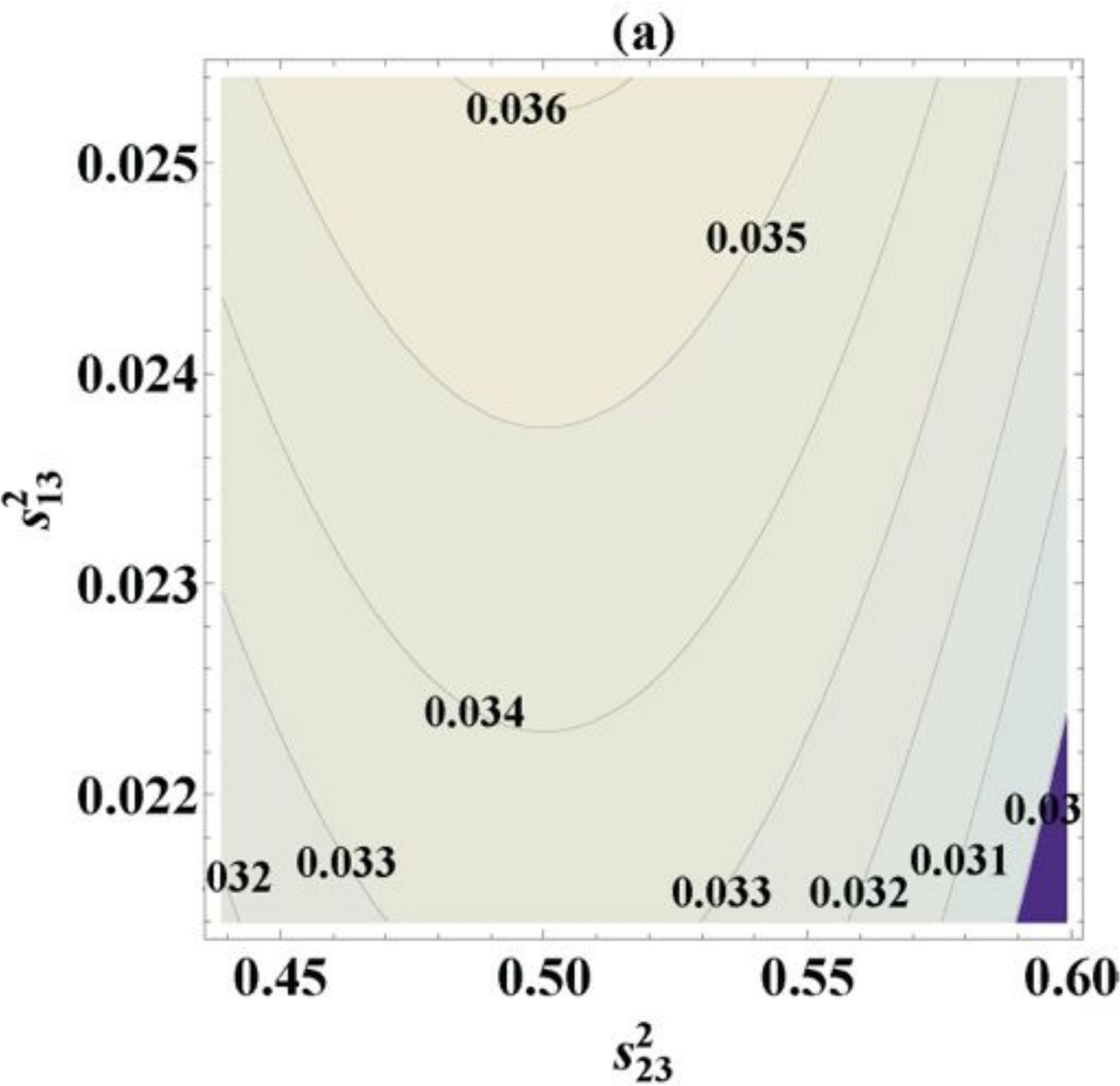}
\includegraphics[width=0.6\linewidth]{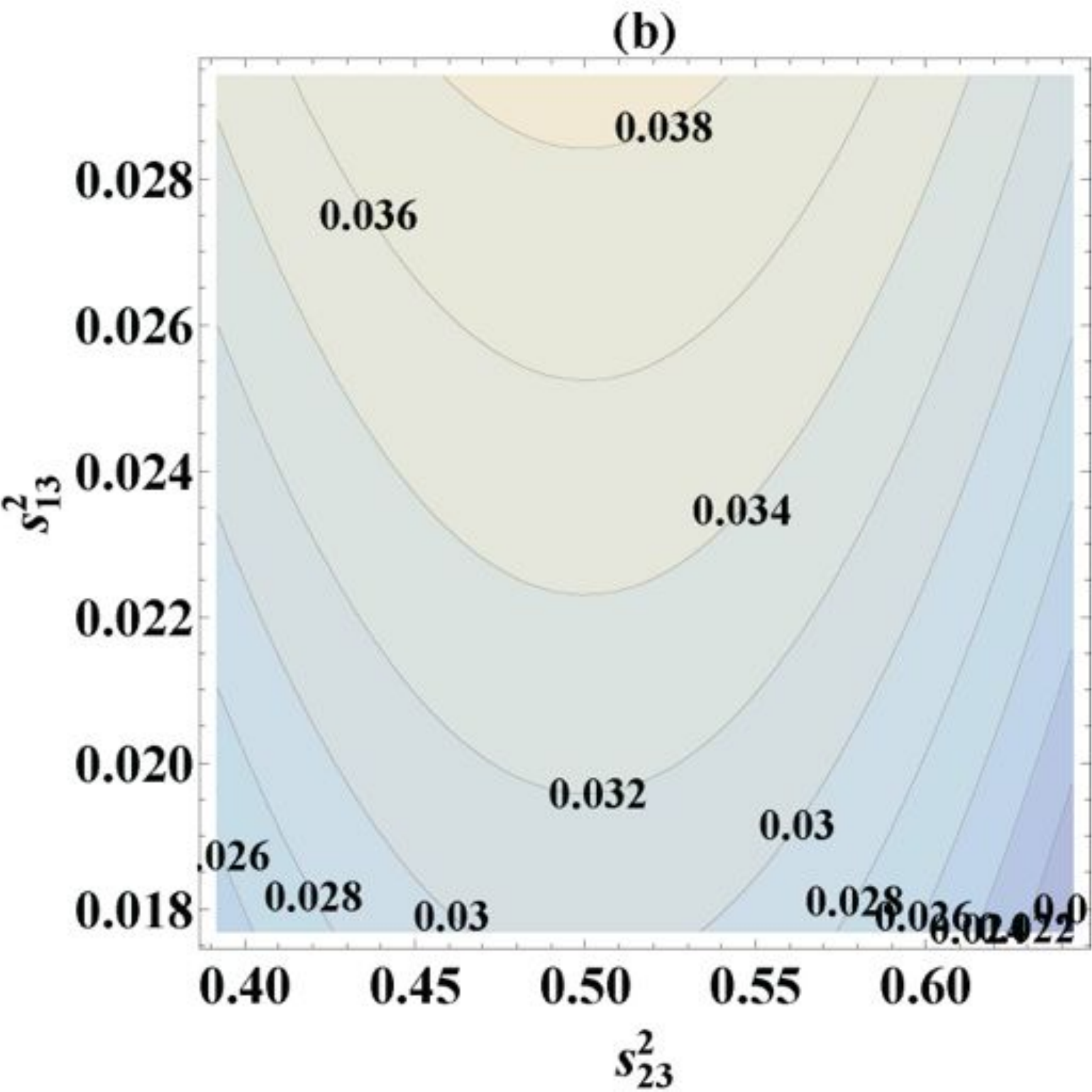}
\end{center}
\caption{Contour plots for each values of ${\rm J_{CP}}$ in the plane ($s_{23}^2$, $s_{13}^2$) for (a) Case A ($1\sigma$),  (b) A ($3\sigma$).}
\label{fig2-1}
\end{figure}

\begin{figure}[h]
\begin{center}
\includegraphics[width=0.6\linewidth]{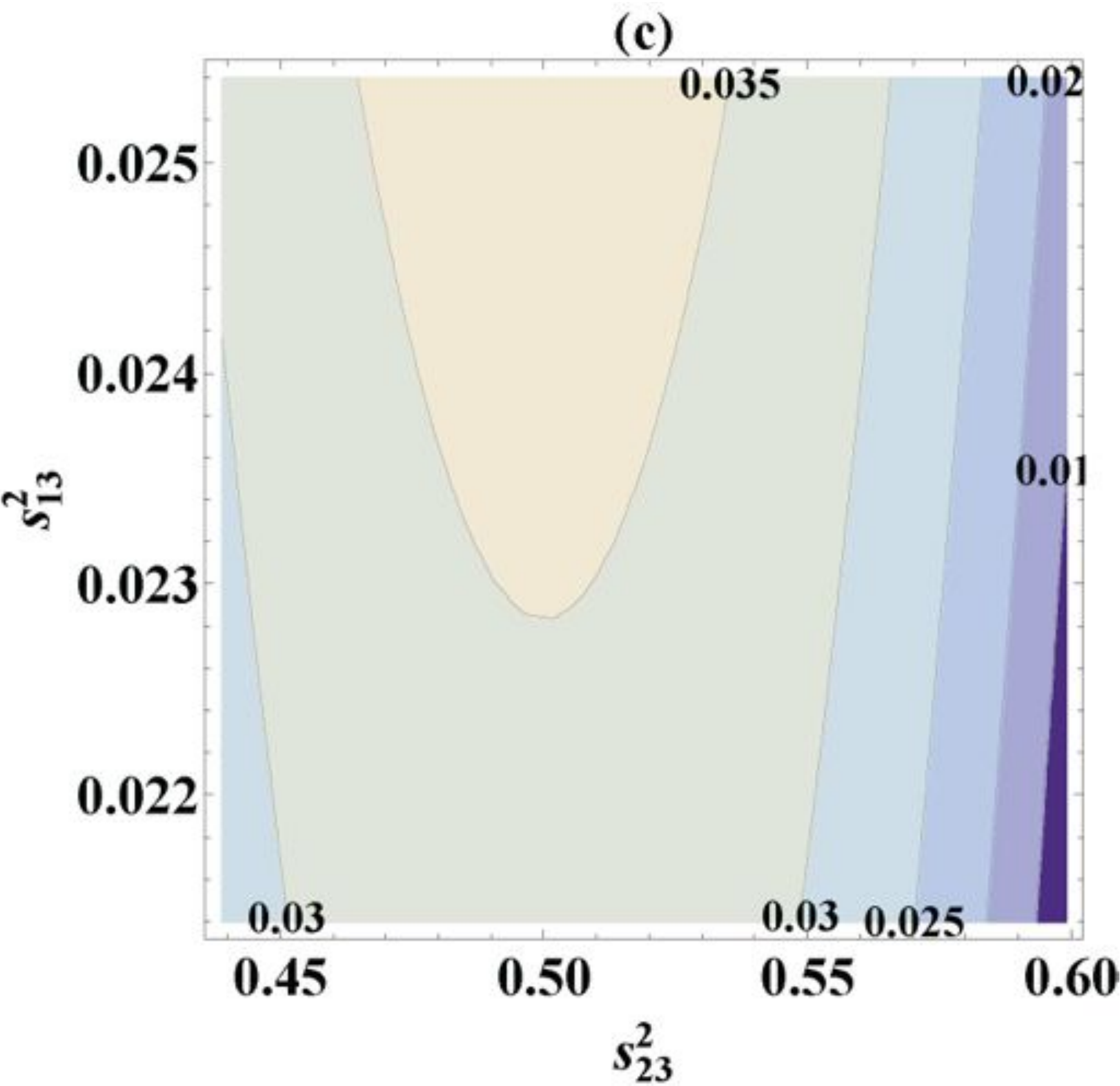}
\includegraphics[width=0.6\linewidth]{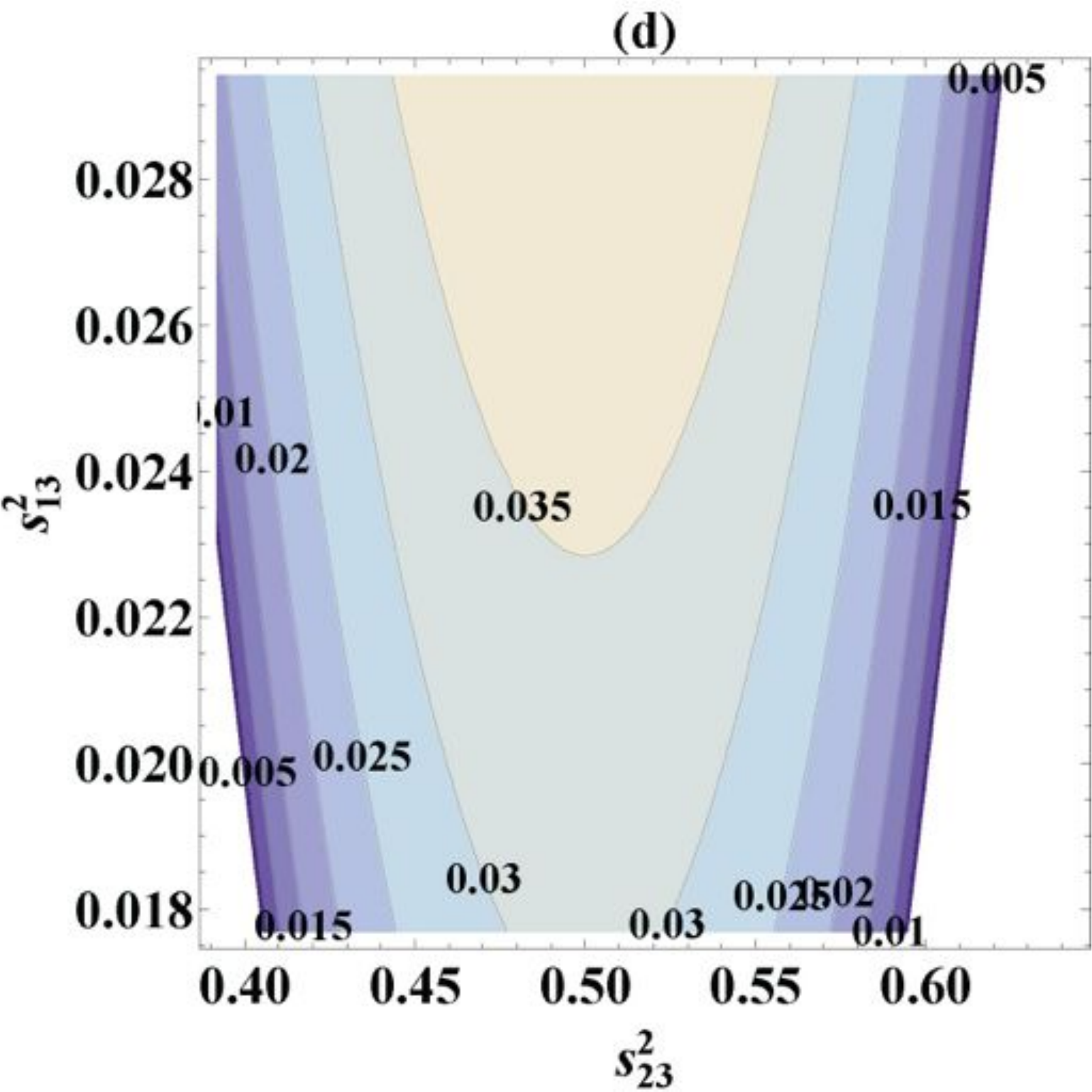}
\end{center}
\caption{Contour plots for each values of ${\rm J_{CP}}$ in the plane ($s_{23}^2$, $s_{13}^2$) for (c) Case B ($1\sigma$) and (d) B ($3\sigma$).}
\label{fig2-2}
\end{figure}

\subsection{Numerical Results}
 For our numerical analysis, we take the current experimental data for three neutrino mixing angles as inputs,
which are given at $1\sigma-3\sigma$ C.L., as presented in Ref. \cite{cp-fit2}.
%Note that our numerical results for the Dirac-type CP phase do not depend on the neutrino masses at all, but  the experimental results for three neutrino mixing %angles depend on the hierarchical structure of neutrino mass spectrum.
Here, we perform numerical analysis and present results only for normal hierarchical neutrino mass spectrum.
It is straightforward to get numerical results for the inverted hierarchical case.
Using experimental results for three neutrino mixing angles, we  estimate  the values of  $\delta_D$ and ${\rm J_{CP}}$ in terms of neutrino mixing angles through Eqs. (\ref{phaseA}) and (\ref{jcp1}), respectively, and the formulae presented
in Table I.

\begin{figure}[h]
\begin{center}
\includegraphics[width=0.6\linewidth]{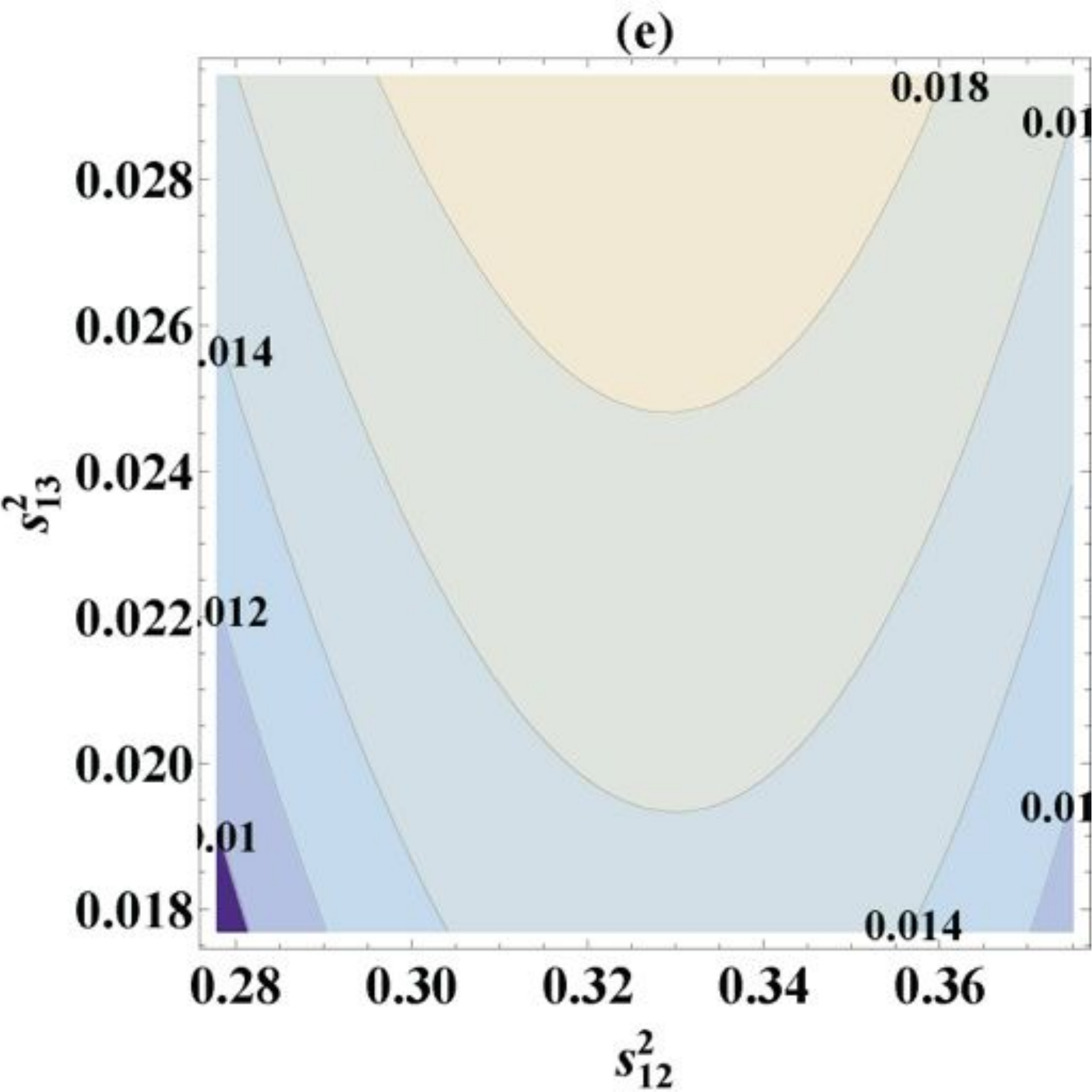}
\includegraphics[width=0.6\linewidth]{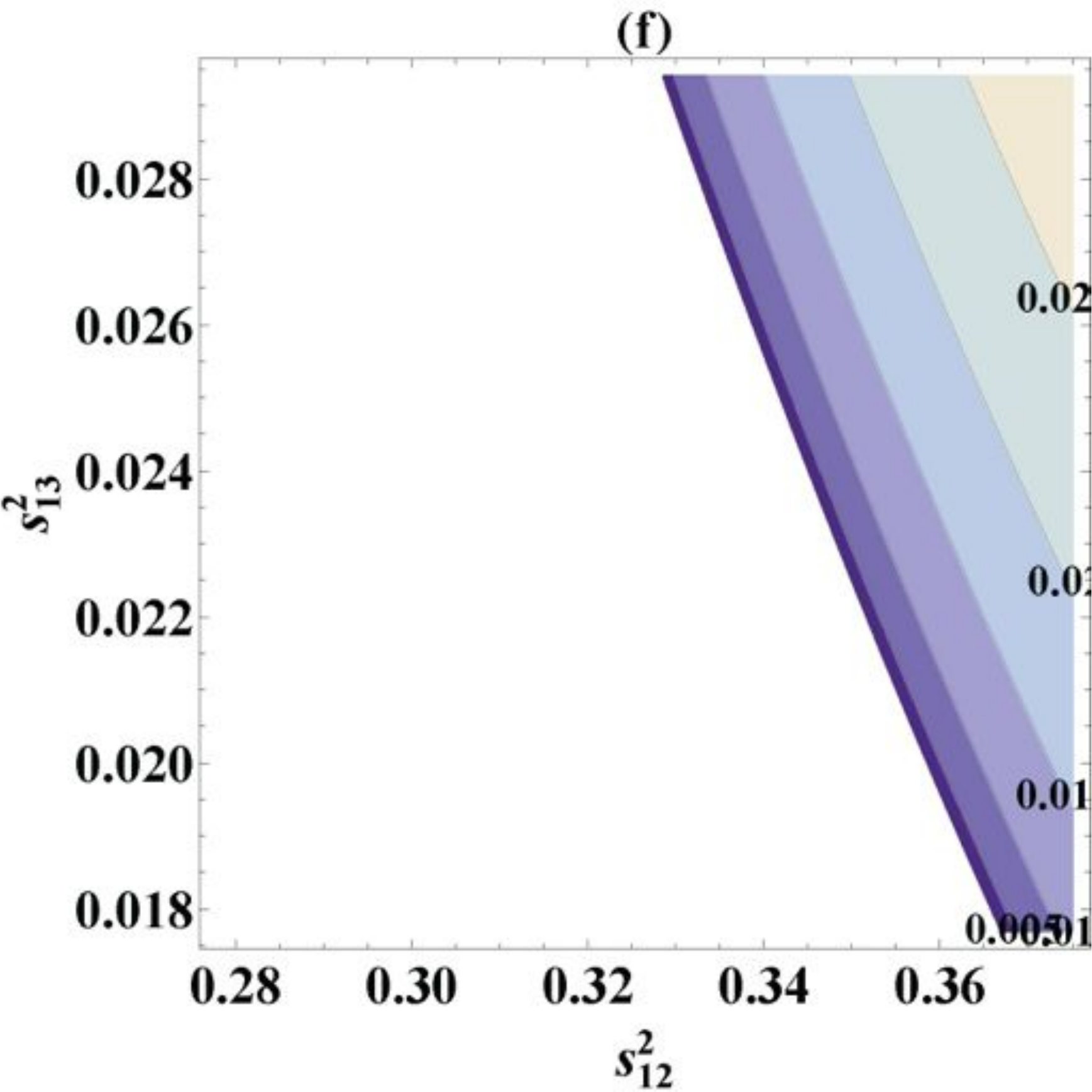}
\end{center}
\caption{Contour plots for each values of ${\rm J_{CP}}$ in the plane  ($s_{12}^2$, $s_{13}^2$) for  Cases  (e) C, D  and (f)  E, F.}
\label{fig3}
\end{figure}

Fig. \ref{fig1} shows the predictions of $\delta_D$ in terms of $s^2_{23}$  [(a): Cases A and B] and $s_{12}^2$ [(b): Cases C and D]  [(c): Cases E and F] based on the corresponding experimental data given at $3\sigma$ C.L.
Regions surrounded by  blue (red) lines correspond to Cases A, C and E (B, D and F).
In particular,  the small dark regions in Fig. \ref{fig1}-(a) and (b) correspond to the results obtained by using the experimental data at $1\sigma$ C.L. for Cases A-D
which apparently indicate CP violation.
The width of each bands implies the variation of the other mixing angles, $s_{12}^2$ (Cases A and B) and $s_{23}^2$ (Cases C-F). We see that  alsmost maximal $\delta_D\sim \pi/2, 3\pi/2$ can be achieved by $s^2_{23}\sim 0.5$  for Cases--A, B and by $s^2_{12}\sim0.325$ for Cases--C,D.
It turns out that  the magnitude of CP violation is not large for Cases E and F.

In Figs. \ref{fig2-1}, \ref{fig2-2} and \ref{fig3}, we display contour plots for each value of  $|{\rm J_{CP}}|$ in the planes of ($s_{23}^2$, $s_{13}^2$) (a-d) and
 ($s_{12}^2$, $s_{13}^2$) (e,f). The panels (a) [(c)] and (b) [(d)] correspond to the results for Case--A(B) obtained by using the experimental data at $1\sigma$  and $3\sigma$ C.L., respectively.
 The results for Cases  C (D) and E (F) based on the experimental data at $3\sigma$ C.L. are displayed in the panels
(e) and (f), respectively.
 We note that the sizes of  $|{\rm J_{CP}}|$ in the lepton sector  for Cases A and B can be as large as $0.03\sim 0.04$ which are much larger than the values of the quark sector  as order of  $10^{-5}$, and expected to be measurable in foreseeable future.
Such a large value of $|{\rm J_{CP}}|$ can be anticipated from Eq.(\ref{jak}) by imposing the experimental values of neutrino mixing angles for large CP phase $\delta_D\sim \pi/2$, since we are led from Eq.(\ref{jak}) to 
${\rm J_{CP}}\sim 0.35 \sin\delta_D$
for the central values of experimental data for the neutrino mixing angles.
For cases C, D, and F, most parameter space predicts the values $|{\rm J_{CP}}|$ less than 0.3,  as shown in Fig, \ref{fig3} (e) and (f).
We see from Fig. \ref{fig3} (f) that the region of $s^2_{12}<0.32$ for Cases E and F is excluded because it  leads to $|\cos\delta_D|>1$ for the experimentally allowed region of $s^2_{13}$.
%\\
%
%\section{Conclusion}
%As a summary, we have considered an Ansatz to estimate possible size of the  Dirac-type CP phase $\delta_D$ with regards to
%two neutrino mixing angles in the standard parametrization of the neutrino mixing matrix.
%This has been achieved by equating  one of {\it  minimally-modified (tri-)bimaximal} (M(T)BM)  parameterizations of the neutrino mixing matrix with the standard parametrization of the PMNS one.
% Through the procedure, we could obtained several relations expressed in terms of two or three neutrino mixing angles and Dirac-type CP phase.
%We have also discussed how the  parameterizations of the neutrino mixing matrix can be related with symmetries associated with tri-bimaximal or bimaximal mixing.
%However, the scheme we proposed could not lead to the predictions on the Majorana phases yet in terms of the mixing angles.
%Carrying out numerical analysis based on the current experimental results for  neutrino mixing angles, we have  obtained the values of the Dirac-type CP phase for several cases.
%\\
%

%%%%%%%%%%%%%%%%%%%%%%%%%%%%%%%%%%%%%%%%%%%%%%%%%%%%%%%%%%%
%\newpage
%%%%%%%%%%%%%%%%%%%%%%%%%%%%%%%%%%%%%%%%%%%%%%%%%%%%%%%%%%%%%%%%%%%%%%%%%%%%%%%%%%%%%%%%%%%%%%%%%%%%%%%%%%%%%%%%
\noindent{\bf Acknowledgments}

The work of C.S.K. was supported by the NRF
grant funded by Korea government of the MEST (No. 2011-0017430) and (No. 2011-0020333). The work of S.K.K. was supported by the NRF
grant funded by Korea government of the MEST (No. 2011-0029758).

%\newpage
%%%%%%%%%%%%%%%%%%%%%%%%%%%%%%%%%%%%%%%%%%%%%%%%%%%%%%%%%%%%%%%%%%%%%%%%%%%%%%%%%%%%%%%%%%%%%%%%%%%%%%%%%%%

\end{document}